# High-Efficiency Lucky Imaging

Craig Mackay

## ABSTRACT

Lucky Imaging is now an established observing procedure that delivers near diffraction-limited images in the visible on ground-based telescopes up to ~2.5 m in diameter. Combined with low order adaptive optics it can deliver resolution several times better than that of the Hubble Space Telescope. Many images are taken at high speed as atmospheric turbulent effects appear static on these short timescales. The sharpest images are selected, shifted and added to give a much higher resolution than is normally possible in ground-based long exposure time observations. The method is relatively inefficient as a significant fraction of the frames are discarded because of their relatively poor quality. This paper shows that a new Lucky Imaging processing method involving selection in Fourier space can substantially improve the selection percentages. The results show that high resolution images with a large isoplanatic patch size may be obtained routinely both with conventional Lucky Imaging and with the new Lucky Fourier method. Other methods of improving the sensitivity of the method to faint reference stars are also described.

## 1 INTRODUCTION

The performance of ground-based telescopes is greatly hampered by the effects of atmospheric turbulence. Even at the best observing sites, the angular resolution is little different from what may be achieved with a 10cm diameter telescope in the visible, typically ~one arcsec. Results from the Hubble Space Telescope (HST) have revolutionised most branches of astronomy by delivering a resolution dramatically better than routinely achieved on ground-based telescopes. Adaptive optics techniques, sometimes together with laser guide stars have produced similar resolution images in the near infrared on larger (8-10m) ground-based telescopes. However, despite the substantial investment in these technologies, adaptive optics has been unable to deliver Hubble resolution in the visible from any ground-based telescope. The only technique to achieve this routinely on faint targets is Lucky Imaging (LI) (Mackay et al. 2004 and Law, Mackay and Baldwin, 2006). LI relies on taking a series of images rapidly enough to freeze the motion due to atmospheric turbulence. The stochastic nature of the turbulence means that in some frames the images will be very sharp, close to diffraction limited. Provided a reference star is in the field of view bright enough to allow image sharpness to be measured, images may be selected and ranked by quality. The best percentage are shifted and added to give a summed image; The highest resolution images are obtained with the smallest percentages. The image in Figure 1 was obtained using the best 20% of images. It is clear that it is comfortably able to equal the resolution of the HST.

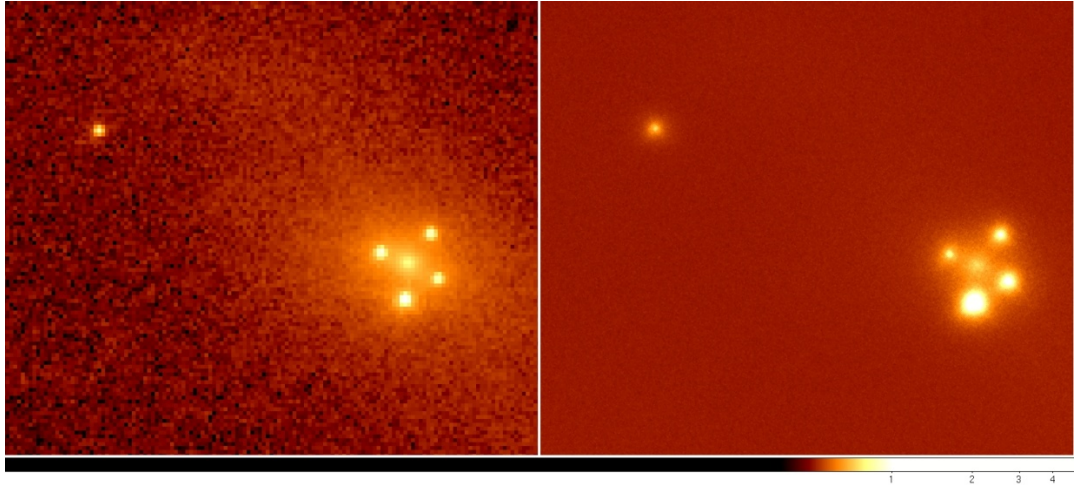

**Figure 1:** Images of the Einstein cross gravitational lens. The light from a distant quasar (redshift z=1.695) is bent by a masive object in the core of a relatively nearby (z=0.0394) Zwicky galaxy ZW 2237+030 seen as the fuzzy object between the four quasar images. The image on the left was taken with the Hubble Space Telescope Advanced Camera for Surveys (ACS) while the one on the right was taken by our Lucky Imaging camera on the NOT telescope on La Palma. The single star at the top left of the image is approximately 9 arcseconds from the centre of the lensing Galaxy. Quasar variability combined with different light travel times for each of the four images changes their apparent relative brightnesses.

Unfortunately, with telescopes significantly larger than 2.5m in diameter, the probability of obtaining an acceptable number of near diffraction limited images becomes very small even on the best ground-based sites. By combining LI with a low order adaptive optics (AO) system, much higher image resolution may be obtained. Using the Palomar 5m telescope, Law et al. (2009) obtained a resolution of 35 milliarcseconds (mas) full width at half maximum (FWHM) at around 770 nm, approximately 20 times better than the prevailing seeing conditions when selecting 5-20% of the images. The images taken are still the highest resolution images ever taken of faint astronomical objects in the visible or near infrared. A high-resolution image of the globular cluster M15 is shown in Figure 2.

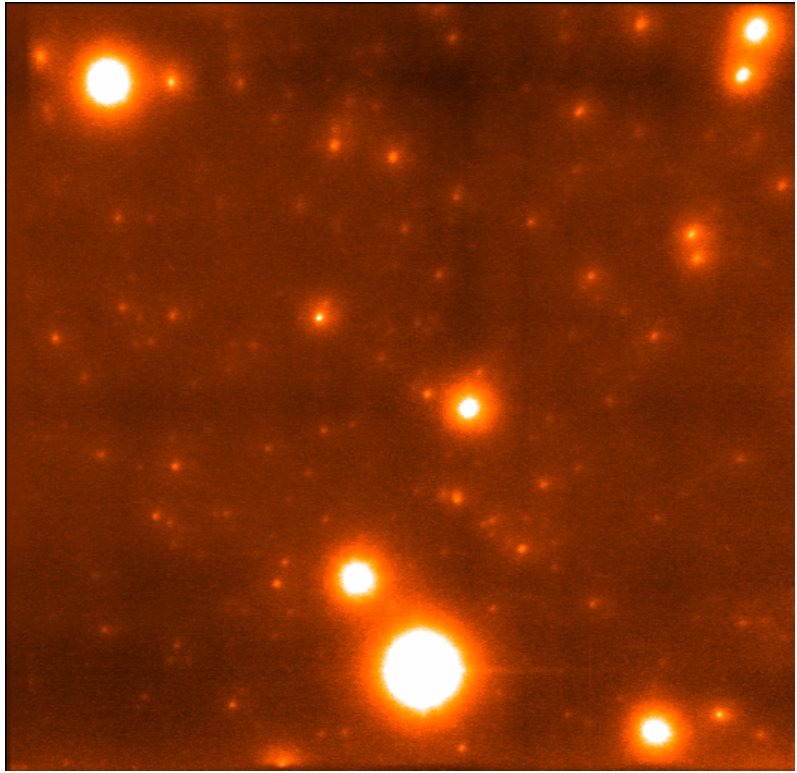

**Figure 2:** The globular cluster M 13 in I band observed with the Lucky Imaging Camera behind the low order PALAO adaptive optic system on the Palomar 5 m telescope. The resolution in this image is about 35 mas or approximately 3 times that of the (undersampled) Hubble Space Telescope. This is the highest resolution image of faint objects ever taken in the visible or infrared anywhere by anyone. This is a 10% Lucky Image selection from the data set.

A detailed study of the images rejected by the percentage selection method shows that very often the resolution is only poor in one direction while other directions are much sharper. The high-resolution information present is potentially very valuable and, if a way could be found to incorporate it in the final images, the resulting signal-to-noise would be greatly improved. Garrel, Guyon & Baudoz (GGB, 2012) have proposed that LI selection be done in Fourier space rather than conventional image space. In their paper, GGB give a detailed account of their method, which we call Lucky Fourier (LF), and describe a series of simulations of the predicted performance of their method on the 8m Subaru telescope under seeing conditions of 0.6 arcseconds. This paper reports results of implementing their algorithms and running these on a range of Lucky Imaging datasets taken over recent years. The datasets include both conventional Lucky Imaging runs on the NOT 2.5m telescope on La Palma as well as datasets taken with the Lucky Imaging camera together with the PALMAO low order adaptive optic system on the Palomar 5m telescope at a variety of instrumental resolutions and fields of view. In every case the results agreed well with the predictions of GGB.

In addition, we will look at some other features of LI that can allow the use of significantly fainter reference stars than previously thought possible.

## 2. CLASSIC LUCKY AND LUCKY FOURIER SELECTION

The classic approach to LI selection (Law, Mackay & Baldwin, 2006) starts with a large number of frames within which there is a reference star of "adequate" signal-to-noise, ideally more than about 140 detected photons per frame. This corresponds to I ~ 16.5m on a 2.5 m telescope. Althought the LI method works with any high-speed imaging camera in practice faint reference stars force the use of efficient photon-counting cameras. All our data were taken with photon-counting EMCCD cameras (Mackay et al., 2012).In each frame the position of the brightest speckle of the reference star is used to establish the shift needed to move each frame into registration and then add it into the total. By selecting a subset of the frames for sharpness on the basis of the brightness of the brightest speckle, the angular resolution may be improved significantly. Even with 100% selection the shift and add procedure essentially eliminates the contribution to image smear caused by tip-tilt due to turbulence or telescope pointing errors. The signal-to-noise of the reference star is important since at low signal levels the brightness of the central speckle can depend as much on the statistics of photon arrival times as it can on the intrinsic phase variance of the incoming wavefront. For critically sampled or undersampled images it is generally necessary to use subsampling so that each frame may be located with an accuracy of a fraction of a pixel so as not to compromise the accumulated angular resolution.

The highest resolution images are obtained by choosing the smallest percentage of the very best frames that have the sharpest point spread functions (PSFs). In many cases it is clear that slightly poorer images are smeared in one direction and yet still have the full resolution in the orthogonal direction. The method proposed by GGB is intended to enable that full resolution information to be used to improve the signal-to-noise while preserving overall angular resolution when more frames are combined. This could help to overcome one of the less satisfactory aspects of the LI method which only delivers high resolution when a large fraction of the images are discarded.

The Fourier transform of a two-dimensional astronomical image has peaks at the four corners. It is easier to follow what is happening in Fourier space if the zero frequency component is moved to the centre of the Fourier array. This is done by moving each quadrant of the 2-D Fourier transform diagonally across the array (this process is reversed before the inverse transform is carried out). The Fourier transform will now have a relatively narrow peak in the middle of the array. The width of that peak is related to the sharpness of the original image with the peak being wider for higher resolution images. If the original image has a circular point spread function then the central peak in the Fourier plane will also be circular. In the case of an image with much better resolution in one direction than another then that will be reflected in the Fourier peak so that it is wider in the direction that corresponds to the high-resolution vector in the image plane.

Classic LI relies on attributing a quality rating to whole images on the basis of the sharpness of the reference star in each frame. In the Fourier plane, the LF method suggested by GGB uses the amplitude of each element in the complex (u,v) plane i.e. components of the image rather than the full image in conventional LI. The percentage selection is then made amongst the corresponding elements in each frame of the full data set. For any particular element in the complex (u,v) plane, the highest amplitude (u,v) elements will be derived from different frames and so each (u,v) location needs to be ordered by amplitude independently. Looking at the 10% selection frame, as an example, we see that each (u,v) complex element is the complex sum of the 10% highest amplitude (u,v) elements in the entire sequence  The 10% selection frame therefore consists of the average of the highest 10% of (u,v) elements recorded for each and every (u,v) location independently.  This 10% Fourier transform is then inverse transformed to give the 10% selected output image.  To see the effect of the method, an LF image may be compared directly with the 10% classic LI image. In practice the images are first processed to accurately align them as in the conventional LI method.  This avoids phase ramps across the Fourier transform that make it harder to follow how the process works.

It is important to address a further complication with the Fourier method.  If we think about what different regions of the Fourier plane represent we see that the central peak gives the information about the point spread function across the image while the outer parts of the Fourier plane provide information about the geometry of the image, essentially where each object in the image is located.  We have observed that one of the particularly valuable properties of LI is that the sharpest images are the ones that are found to have the largest isoplanatic patch size.  This makes LI a particularly valuable technique for delivering high resolution images over a good sized field of view.  When selecting images in Fourier space rather than in the image space, we are likely to compromise that geometry.  To minimise this effect, the selection in Fourier space is done as described above within a definedlow spatial frequency region of the Fourier plane.  Outside this region, the selection is made in image space.  This is done by complex summing the (u,v) elements of the sharpest 10% of images as determined conventionally in image space while the complex summing of the (u, v) elements inside this low spatial frequency region is done on the highest amplitude 10% of the (u, v) elements.  This hybrid summed Fourier image, with the low spatial frequency parts chosen according to the GGB procedure and the high spatial frequency parts chosen essentially from the standard lucky imaging procedure, is then Fourier converted to give the output image.  There is a further reason to restrict the Fourier selection to a low spatial frequency patch in the Fourier plane.  If we think not about the best images but the very worst ones then these are images where each compact object appears broken up into multiple diffraction-limited speckles.  Such an image contains significant high resolution power but the overall image geometry and the isoplanatic patch size will be very poor.  Again, using data over the whole Fourier plane from such an image would substantially compromise the overall image quality.  As it is always difficult to be clear exactly what is going on in Fourier space it is essential to check that the output images are consistent with conventional lucky by checking that the fluxes and astrometric positions of the objects in the field are properly preserved.

## 3.  IMAGE SHARPNESS AND STREHL RATIOS

The Strehl ratio of a star image is defined as the peak brightness of that image divided by the peak brightness that would be achieved with a system that is optically perfect,in the absence of atmospheric effects.  In practice, astronomical images are rather complex and a considerable amount has already been written about the nature of the seeing profile on real, ground-based telescopes (e.g. Trujillo et al. (2001) and references therein).  The point spread function (PSF)is typically described as having a Gaussian core with a more extended outer profile best described by a Moffat function (Moffat, 1969).  These "wings" are broadly predicted by atmospheric turbulence theory although their extent and the amount of flux within them depends on the exact conditions. The Strehl ratio is a parameter often used by optical engineers to describe the quality of design. In some astronomical instances, achieving a high Strehl ratio is very important. For example the detection of faint planets around a bright star requires exceptional optical performance. However for the great majority of astronomical applications high Strehl ratio is relatively unimportant. With typical seeing of ~1 arcsec, star images often have a Strehl ratio of less than 1%. Despite that, an extraordinary amount of astronomy has been done successfully. LI may also be judged by the delivered Strehl ratio which can be in excess of 25% on a 2.5 meter telescopes under reasonably good conditions. LI also gives point spread functions that are well described by a Moffat function. However it has been demonstrated that the various estimates of Strehl ratio give rather different results particularly if the data are sampled close to the Nyquist limit. In practice, a more useful measurement of image quality is to look at the light concentrated within a particular area. In this paper we look at the fraction of light in the star image inside a circle of diameter half the diameter of the first Airy zero (at 1.22 $\lambda$/D, where $\lambda$ is the wavelength and D is the telescope diameter)

## 4. TRIAL DATASETS

Classic LI has produced a large number of important results using data taken on telescopes up to around 2.5 m in diameter. For these tests we have used data taken on the Nordic Optical Telescope (NOT) on La Palma, a 2.5 m telescope on an excellent site. The data were taken using a quad-EMCCD photon counting camera developed in Cambridge with 1024 x 4096 pixels. For simplicity we have restricted our trials to a single 1024 x 1024 pixel area in the field of view. The data used had already been corrected for any instrumental defects such as bias uniformity and cosmic ray contamination. The EMCCD is operated as a high gain photon counting camera and the data were thresholded to allow the identification of individual photons. This produced the highest detection efficiency and a reduced background noise level. The frames were recorded at a rate of 21 Hz. Images of the core of the globular cluster M15 were recorded on 23 July 2009 in SDSS I band (770 nm central wavelength, 160 nm pass band). The field was chosen because it has a large number of stars within the 1024 x 1024 field of view, so allowing detailed studies of the isoplanatic patch size. The zenith distance of the observations was 21° and the natural seeing as measured by a nearby Differential Image Motion Monitor (DIMM) was 0.73 arcseconds, close to the median seeing value for the site. At that wavelength the diffraction limit of the telescope, taken to be 1.22 $\lambda$/D is about 75 mas. The 13 $\mu^2$ pixels of the EMCCD detector correspond to 32.5 mas square so that the PSF is well sampled. TtThe data were sub sampled onto a 4x4 grid so that each pixel was then 8.125 mas square.

Classic LI is only practical on telescopes up to 2.5 m diameter even on the best astronomical sites. High-resolution imaging has been demonstrated to be possible when LI is combined with low order AO correction. Results obtained on the Palomar 5 m telescope combined with the Cambridge Lucky Camera have already demonstrated the highest resolution images ever taken on faint objects in the visible or near infrared (Law et al. 2009). Images with near diffraction limited cores with 35 mas FWHM were obtained with Strehl ratios in the range of 5%-20% in I band. The data for these runs were obtained in July 2007 with an earlier version of the Cambridge Lucky Camera which provided 528 x 512 pixels read out at 20 Hz frame rate. The Palomar 5 m low order adaptive optics system (PALMAO, see Law et al. 2009) was updated at around 200 Hz during these observations. Several runs have been analysed, with the core of the globular cluster M 13 being observed in SDSS I band with a scale of 20.7 mas per pixel. The data were recorded at 20 Hz frame rates, with 512 x 528 pixel images, giving a field of view of 10.6 x 11 arcsecs. As with the other runs the data were first corrected for instrumental deficiencies such as uniformity and cosmic ray removal, and the data were thresholded to enable photon counting to be used.

## 5. DATA ANALYSIS

With classic LI and a faint reference star, care must be taken in interpreting the apparent sharpness achieved. The position and intensity of the central speckle from the reference star may be affected significantly by photon shot noise statistics. The effect can artificially increase the reference star sharpness and so image quality comparisons involving the reference star can be unreliable particularly if the reference star is faint. In all the datasets analysed here, the reference star is relatively bright and so this effect should be negligible. It is worth noting that in the case of the LF it is the entire image that is Fourier transformed. The reference star is used to measure the image offset and the image translated so that all the images are perfectly aligned. If this is not done then different offsets translate into phase gradients across the Fourier plane. It is worth noting that by Fourier transforming the whole image, the light from all the stars in the field are used simultaneously to measure the image quality even if those stars are too faint individually to be used for classic LI reference.

The analysis starts with data that have been pre-processed through the LI data processing pipeline developed by Staley (Ph.D. thesis, University of Cambridge, 2012). This pipeline removes instrumental artefacts and provides basic calibration. It also computes the best estimate of the position of the reference star in the image. The following steps were implemented using Matlab for speed of development:

1. Each image is expanded and interpolated by 4x4 using a drizzle type interpolation (Fruchter & Hook, 2009).
2. The position of the reference star is used to translate the image to align it with a specified reference position. This ensures that the Fourier transformed image does not have any significant phase gradient across it.
3. In order to minimise unnecessary ringing throughout the Fourier transform process the edges of each image are smoothed off with a simple filter so the amplitude of the image goes linearly to zero over 32 pixels.
4. The registered image is Fourier transformed with double precision arithmetic, and each Fourier transformed image is stored andthe central parts of each one are used to give a three-dimensional data array of (frame number, u,v).
5. A central low spatial frequency patch diameter is selected. Within this patch the Fourier components are ordered by amplitude and outside of this they are ordered in the same sequence determined by conventional LI. The size of the central low spatial frequency patch is selected by trial and error in order to minimise the ringing and background noise level. In practice this was limited to some degree by the available computing systems. The size of this circular patch was typically 1/8-1/16 of the size of the entire Fourier plane.

6. The top percentage (where percentage varies from 1-100%) are vector averaged to give each of the (u,v) values for each of the percentage bins.
7. In order to minimise ringing the transition between the inside and outside of the patch is smoothed over 32 pixels linearly.
8. The output percentage Fourier images are then inverse transformed into image space.

These images are the final LF output images which are used for comparison with conventional LI images from the same original dataset.

**5.1 M 15 Data from the Not Telescope**

The most obvious way to see the improvement in the image quality is to simply look at the images visually.

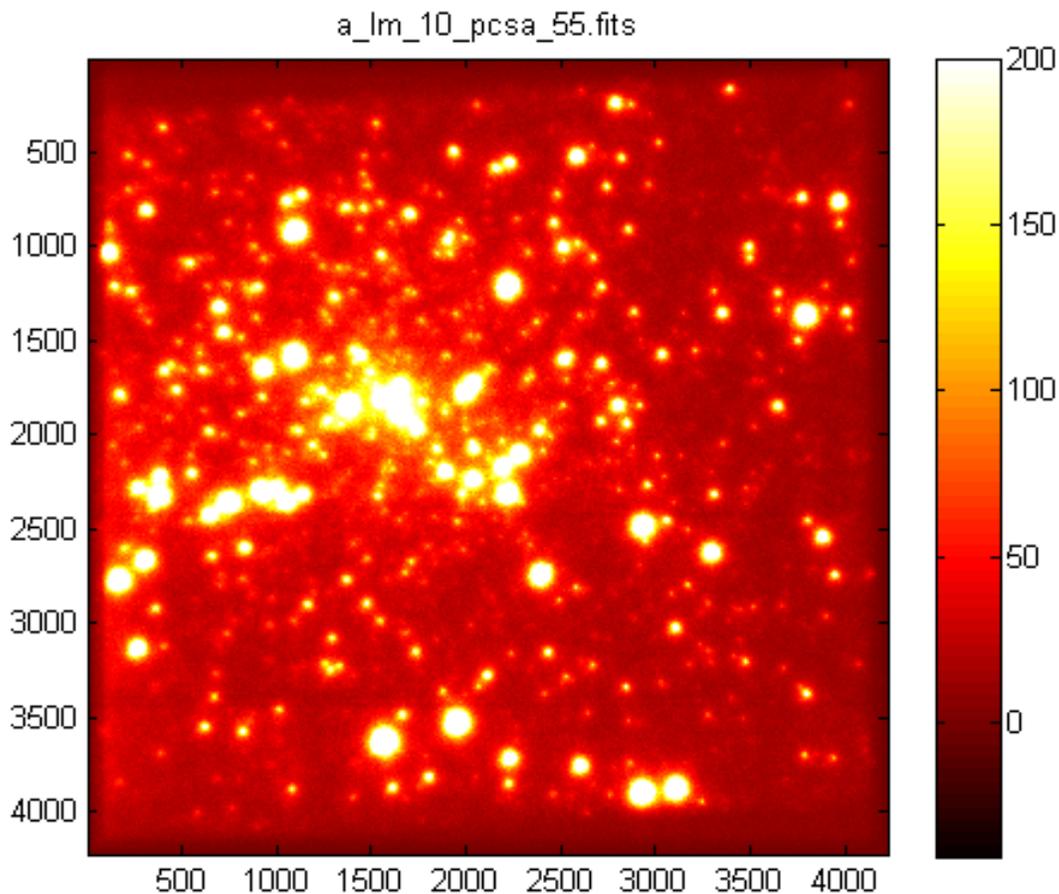

**Figure 3**: A central portion of the globular cluster M 15 reduced with the standard LI method. The field of view shown is approximately 34 x 34 arcseconds (8.1 mas pixels) and the resolution varies from about 0.1 arcseconds close to the reference star which is at approximately (x,y) = (3800, 1400), to about 0.13 arcseconds 40 arcseconds from the reference star. It is apparent visually that the resolution even at a distance is still very good and dramatically better than the 0.73 arcseconds seeing reported by the DIMM monitor for this run.

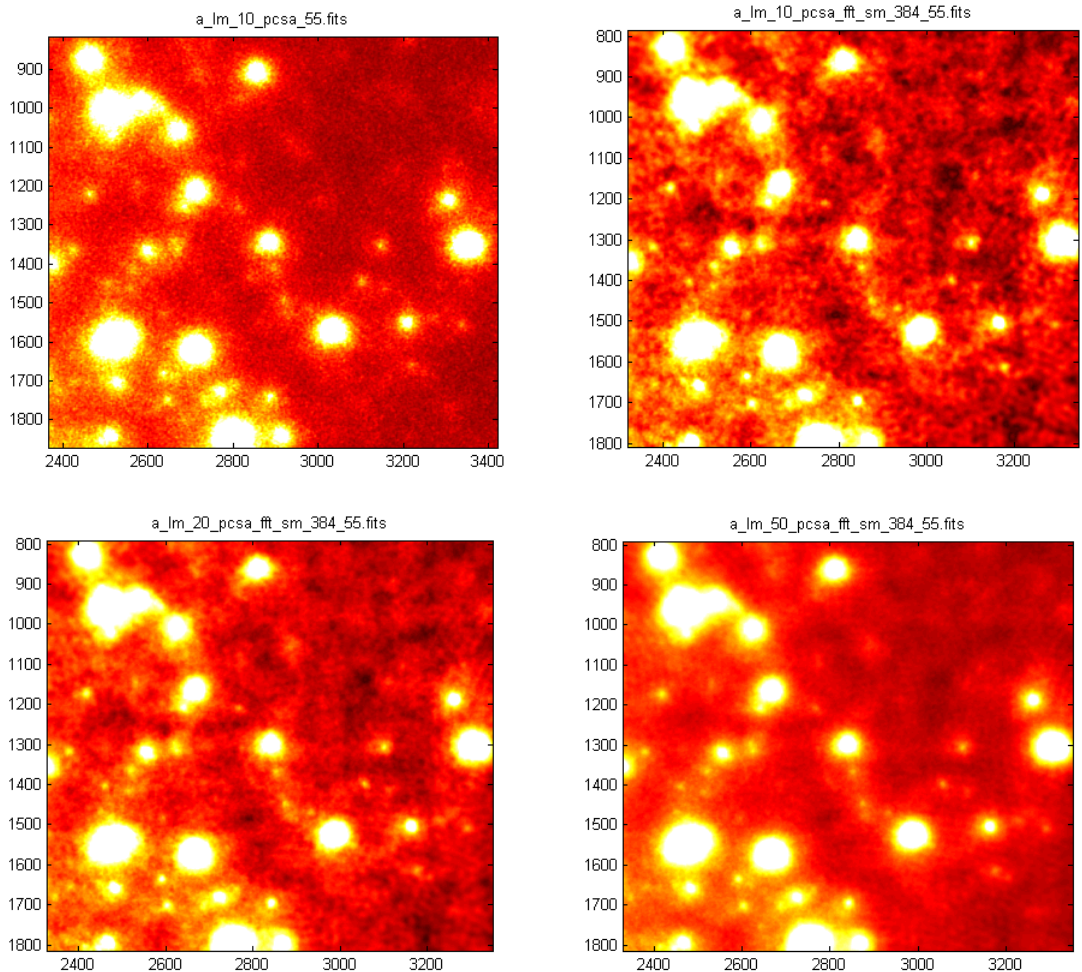

Figure 4: A small section of the field shown in figure 3 approximately 8 x 8 arcseconds. The top left-hand image was obtained using conventional LI selection of the best 10% of the images. The top right-hand image shows the output of the LF procedure also using the best 10% of the images. The bottom left and bottom right images show LF but with 20% and 50% selection. The field shown here is approximately 8 arcseconds from the reference star. The background noise is significantly higher on the Fourier images mainly because of using a rather simplistic apodising procedure limited by our computer capability. It is clear, however, that the sensitivity and image sharpness is very much better using LF even with very high selection fractions such as 50%. All the above images use the same colour look-up table: no rescaling has been applied so they are all directly comparable.

Results may be described more quantitatively. The simplest way to demonstrate the relative improvement in PSF and image sharpness is shown in the figures below. Each profile on the left-hand side is derived from images created using classic LI procedures. The right hand side profiles are derived from the new LF method. In each picture there are 8 curves corresponding with images constructed with 1%, 3%, 5%, 10%, 20%, 30%, 50%, and 100% selection, with the uppermost one corresponding to 1% and the bottom one to 100%. Although with faint targets we would avoid using the reference star to demonstrate how these methods work, in all the cases presented here the reference star is very bright and therefore should be representative of the efficacy of the method. With the LF method, the concept of a reference star does not have any great importance, except that it is used to offset each image before Fourier transforming to minimise any mean phase gradient in Fourier space.

In Fig. 5 we show 4 plots, one of the reference star and then stars at angular separations from the reference star of 11 arcseconds, 23 arcseconds and 31 arcseconds respectively. In every case there is a substantial increase in the both the peak brightness and a corresponding reduction in full width at half maximum using LF particularly for stars near the edge of the field of view. Note that the ringing in the outer parts of the profiles from the LF images is caused by inadequacies in the apodising used in carrying out the Fourier processing.

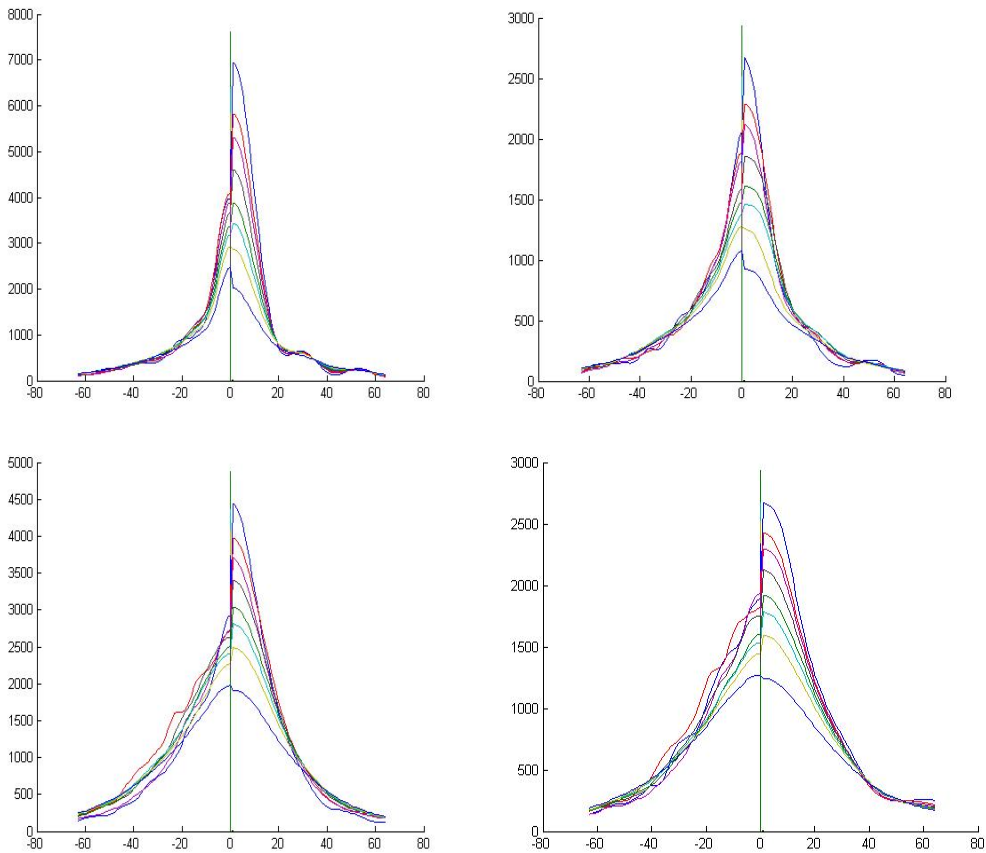

Figure 5: Profiles through four stars in the centre of the globular cluster M15 observed on the Nordic Optical Telescope. The left-hand part of the profile is from the star with conventional lucky imaging and the right-hand half is with the Fourier method. In each image there are eight lines corresponding to 1%, 3%, 5%, 10%, 20%, 30%, 50%, and 100% selection from the top to bottom. The top left-hand star is the reference star, the top right-hand star is at a distance of 11 arcseconds from the reference star, the bottom left star is at 23 arcseconds and the bottom right star is at 31 arcseconds distance. The horizontal scale is in 8.1 mas pixels.

Fig. 6 demonstrates the enhancement delivered by LF when compared with the conventional LI approach.  Here we compare data from the same field processed in the two different ways.  The integrated star fluxes are virtually identical however the flux contained within the diameter of about 75 mas changes as shown.  The one-dimensional plots in Fig. 5 show how the flux is drawn into the centre of the profile by this method.

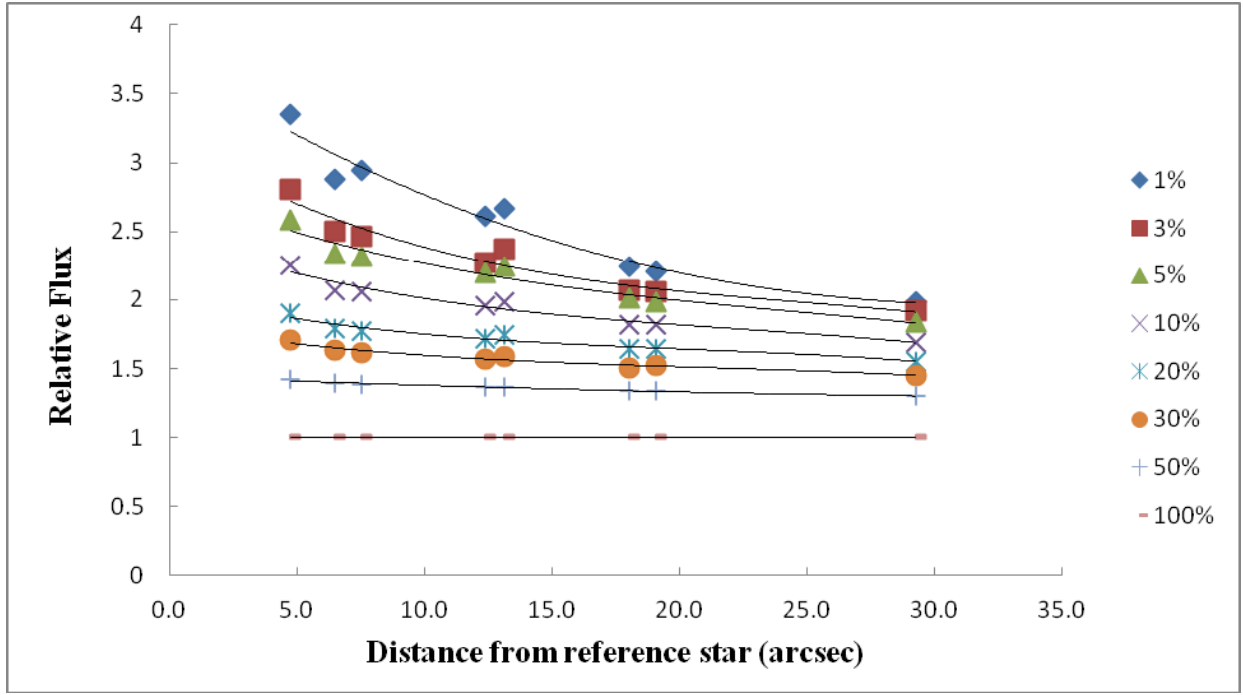

Figure 6: The relative flux from one star measured within the central core of the star image for a variety of different reference stars over distances between 5 and 30 arcseconds from the star.  The relative flux shown here is the value from the LF images relative to the value from the conventional LI data set.

We use a measure of the size of the isoplanatic patch in an image as being the radius at which the Strehl ratio of the reference star is reduced to 1/e or about 37% of its peak value.  Because of the difficulties in measuring a true Strehl ratio in such a crowded field, we measure the core brightness within an area equivalent to the half width of the diffraction limited PSF, about 75 mas in diameter.  As mentioned above, it is not easy to measure image sharpness in the fields that we have chosen as they are full of stars.  Indeed faint stars in the cluster contribute a significant and variable background luminosity.  Rather than try to measure the sharpness of different stars at different distances from a single reference star we reverse the problem and measure the sharpness of one particular star using LI and LF images generated by using reference stars over a range of distances.  This means that any background objects that might affect the measurement of sharpness of the measured star will be present in every instance.  In fields where there are enough stars to do this we believe that this gives a much more reliable value from the true size of the isoplanatic patch.  These data are shown in Fig. 7.

What we see is that the total flux integrated within the core 75 mas diameter circle drops to about 44 % of its value between 5 and 30 arcseconds from the reference star with 1% selection, 55 % with 10% selection, 63% with 30% selection and 67% with 50% selection.  These values indicate for this relatively typical observing run with average seeing  the isoplanatic patch size as defined above is approaching one arcmin in radius or two arcmin in diameter.  This confirms one of the most important features of LI, a very large isoplanatic patch size, and that the LF method in no way compromises this.  The conventional size in the visible on a good site using standard adaptive optics approaches gives an isoplanatic patch size of only a few arcseconds in diameter (Sarazin & Tokovinin, 2001).  However it must be remembered that the definition of isoplanatic patch size used by Sarazin & Tokovinin as related to the phase variance in the wavefront is not what we are using here since we do not have such a measurement.  We simply are looking at the sharpness of the image and the way it declines with distance from the reference object.  The isoplanatic patch size , however it is defined, is a critical parameter because in most astronomical observations we need to be able to compare objects which are often quite far apart when they are faint and we need to have a well defined and understood point spread function that varies across the field in a highly predictable and repeatable way.

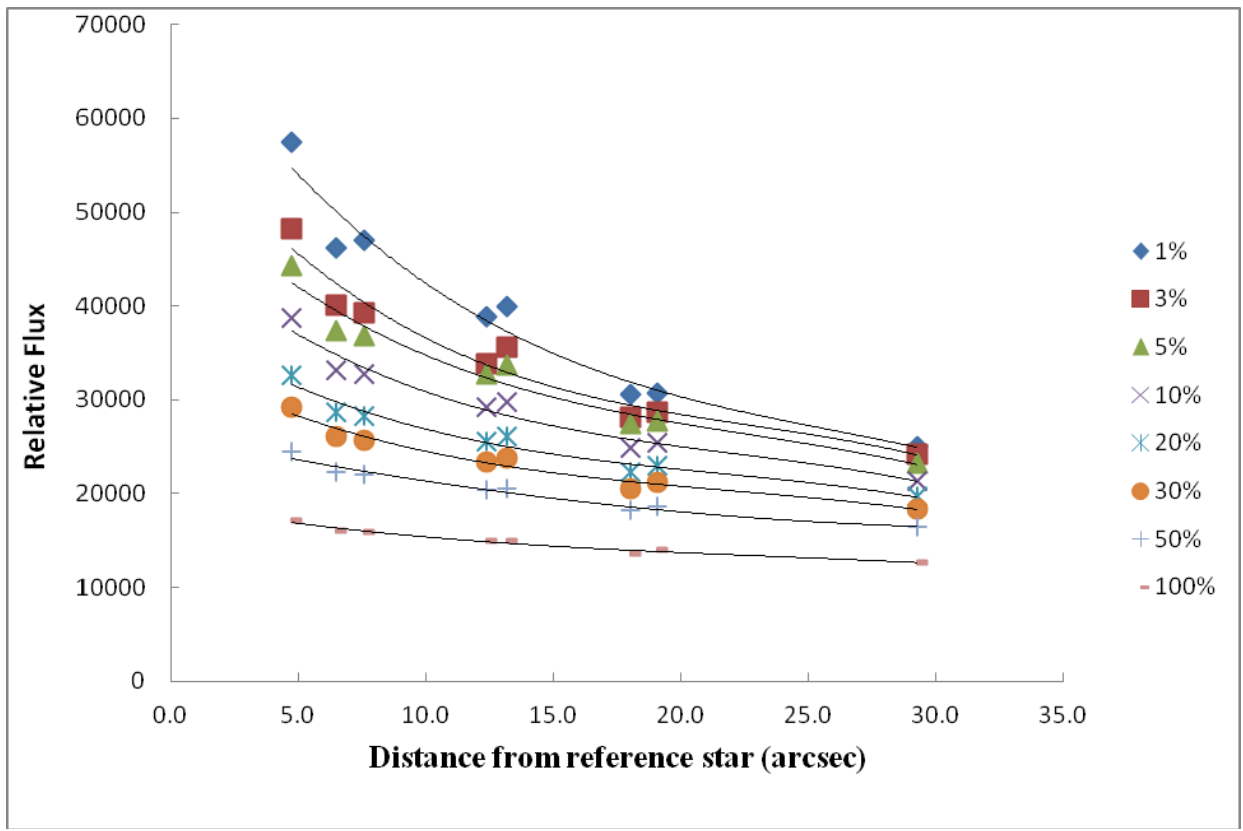

Figure 7: This shows the total flux within the core (~75 mas) of one star using different reference stars over a range of distances between five and 30 arcsec.  Similar results are found with fluxes measured in different ways such as peak brightness, encircled energy etc.

## 5.2 M 13 Intermediate Resolution Data from LuckyCam/PALMAO

We have experimented with a very different dataset taken with the Lucky Camera behind the Palomar low order AO system PALMAO (Law et al. 2009). The data were reduced in the conventional way by performing the LI and LF selection methods described above. The improvement in the images is quite dramatic and can be seen below in Figure 8.

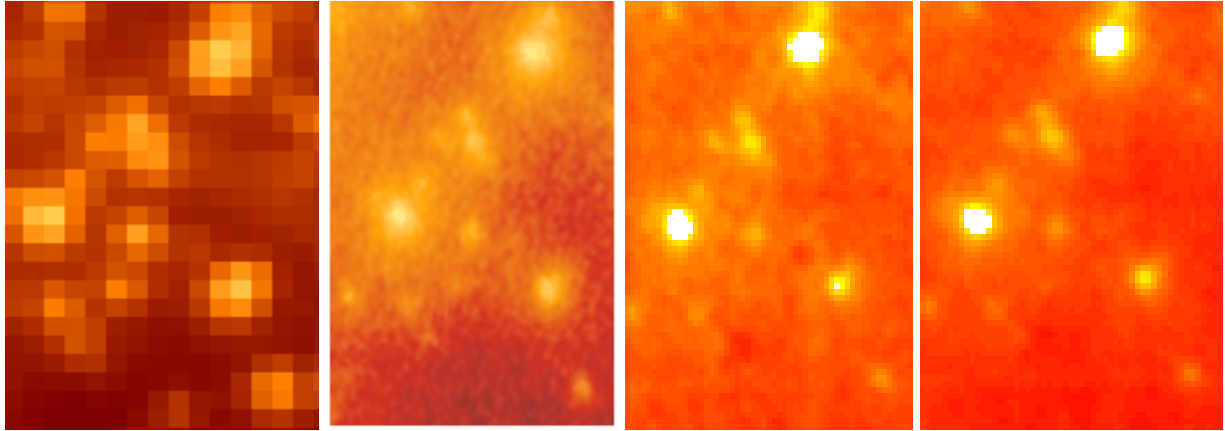

Figure 8: A central portion of the globular cluster M 13 image shown in Fig. 2. The image on the left is from the Hubble Space Telescope Advanced Camera for Surveys, the next one is part of the image shown in Fig. 2 above with 10% LI selection. The next two are using the LF method described above with selection percentages of 20% and 50% respectively. The LF images are significantly sharper and brighter and the diffuse halo around the brightest stars in the image selected conventionally have been significantly reduced in the Fourier images.

It is clear from these images that the resolution obtained with the 50% selection LF method is very similar to that obtained with the conventional image plane selection but at 10% selection. Again this may be seen by looking at the one-dimensional profiles through a small selection of field stars to show the substantial enhancement in angular resolution and central brightness.

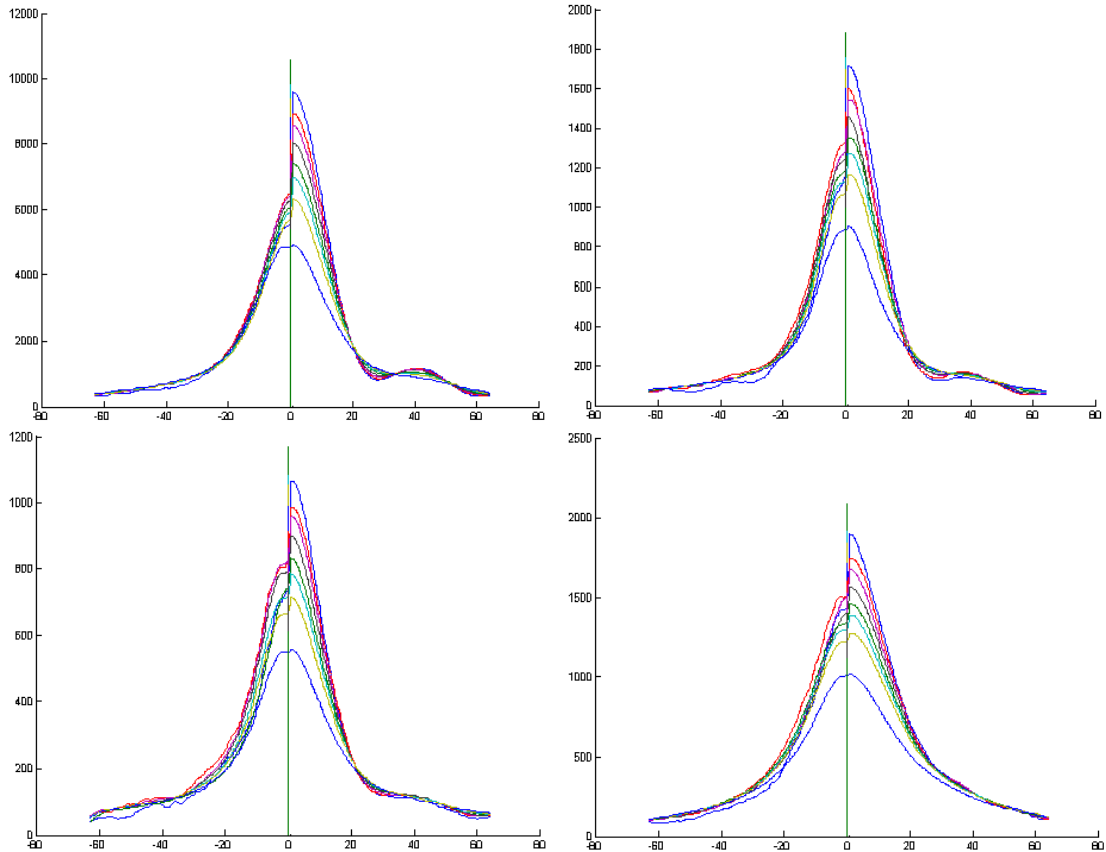

Figure 9: One-dimensionally profiles through four stars in the LuckyCam/PALMAO field. The left-hand part of the profile is from an LI image and the right-hand half is with the LF method. In each image there are eight curves corresponding to 1%, 3%, 5%, 10%, 20%, 30%, 50%, and 100% selection from the top to bottom. The top left-hand star is the reference star, the top right-hand star is at a distance of 1.55 arcseconds from the reference star, the bottom left star is at 3.34 arcseconds and the bottom right star is at 9.2 arcseconds distance. The field of view of this dataset is small, about 10 x 10 arcseconds and the scale in these plots is approximately 5 mas per pixel. The sharpness of the central peak and the width of the profile is markedly improved by the LF method. The width of the star profile increases significantly with distance from the reference star. There are not enough stars in the field to allow an accurate assessment of the isoplanatic patch size but the indications are that it is at least 30 arcseconds in radius. Note that the outer parts of the profile show rings which are not the Airy rings but are an artefact of the way the Fourier processing of the data was done.

## 6. WORKING WITH FAINT STARS IN LUCKY IMAGING

Bright reference objects are very scarce, particularly at high Galactic latitudes, with counts of stars being presented by Simons (1995) in R-band. Using a mean colour index of R-I of 1.5 mags, stars with I=20.5 are found at around one per square arc minute at high Galactic latitudes, I=19.3 at 60°, and I=18 at 40°. The faintest reference stars that can be used for LI and LF determine the fraction of sky that may be covered using these methods at higher Galactic latitudes where stars are scarce. Indeed, at I=20, galaxies are more common than stars at the Galactic poles (Pozzetti et al.). When considering reference objects at faint magnitudes, it is important to remember that many galaxies have compact cores which, if small enough, could also be used as reference objects.

Detailed analysis of many observing runs on the 2.56 meter Nordic Optical Telescope have shown that with guide stars as faint as I = 16 it is possible to get full performance (Law et al. 2006). If there are multiple stars within the field then it is possible to combine the data from them to be satisfactory even if they are individually too faint for full correction. With fainter reference stars such as at I>16.5, the derived sharpness is reduced significantly when compared with that achieved with a much brighter reference star. This is principally due to the effect of Poisson shot noise statistics particularly on the tip-tilt position of each frame which is removed in both LI and LF processing. If the central speckle contains only a few tens of photons then some images will have their positions incorrectly evaluated. This leads to errors in the derived position of each frame which is used to give the alignment before combining the images. However if we look at the derived positions from a typical run there is relatively good correlation over a few frames in these positions. This can be understood since the principal timescale on which this tip-tilt component changes is a fraction of the wind crossing time of the telescope. On La Palma, median wind speeds in the region of 8 m/s give a decorrelation time of a fraction of a second. This is shown in Fig. 10, where the autocorrelation function of the X and Y coordinates of a typical run are shown.

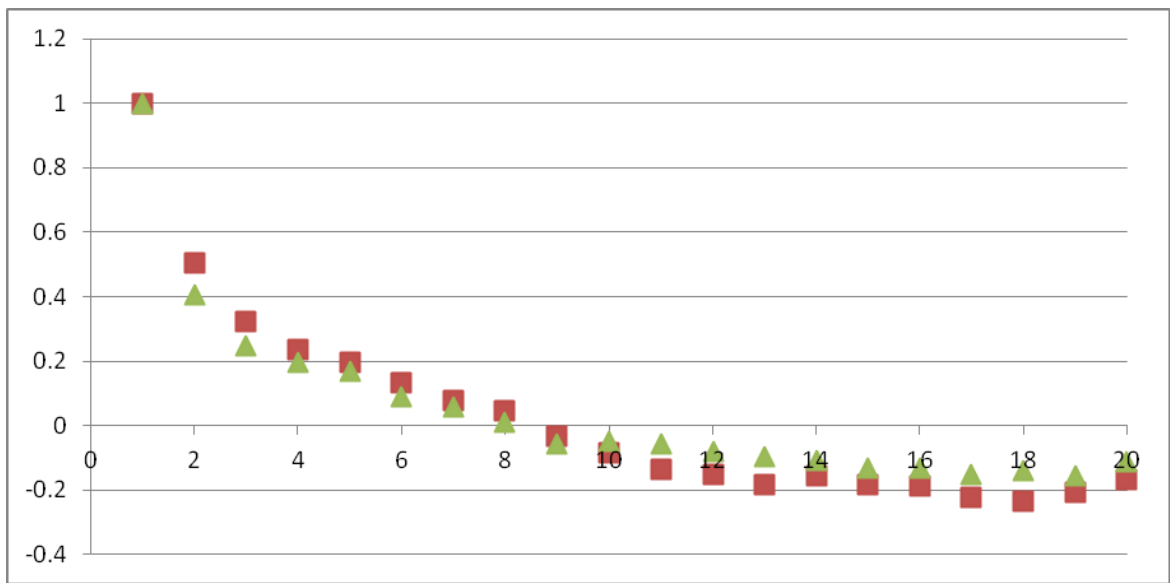

Figure 10. The autocorrelation of the two coordinates describing the position of the peak of the brightest speckle of the reference star on each frame in the sequence used in section 5.1 taken on the NOT. The horizontal scale shows the frame number. Frames were taken at approximately 21 Hz.

We can see that the positions derived in X and Y are quite strongly correlated over a few frames. The correlation essentially vanishes after around 7 or 8 frames. However over a small number of frames the correlation is really quite good. Doing tests to average the X and Y coordinates over a few frames and using very faint reference stars shows that there can be a marked improvement in the quality (as defined by Strehl ratio or core sharpness) over the quality obtained using each frame independently. The effect of this is to reduce significantly the degradation in Strehl ratio because of the lack of brightness of the reference star and therefore enable somewhat fainter reference stars to be used. Our experiments suggest that under a range of relatively good observing conditions it is possible to work with reference stars of 0.5-1 magnitude fainter than the limit described above. This nearly doubles the fraction of sky that lucky imaging may be applied to at high Galactic latitudes.

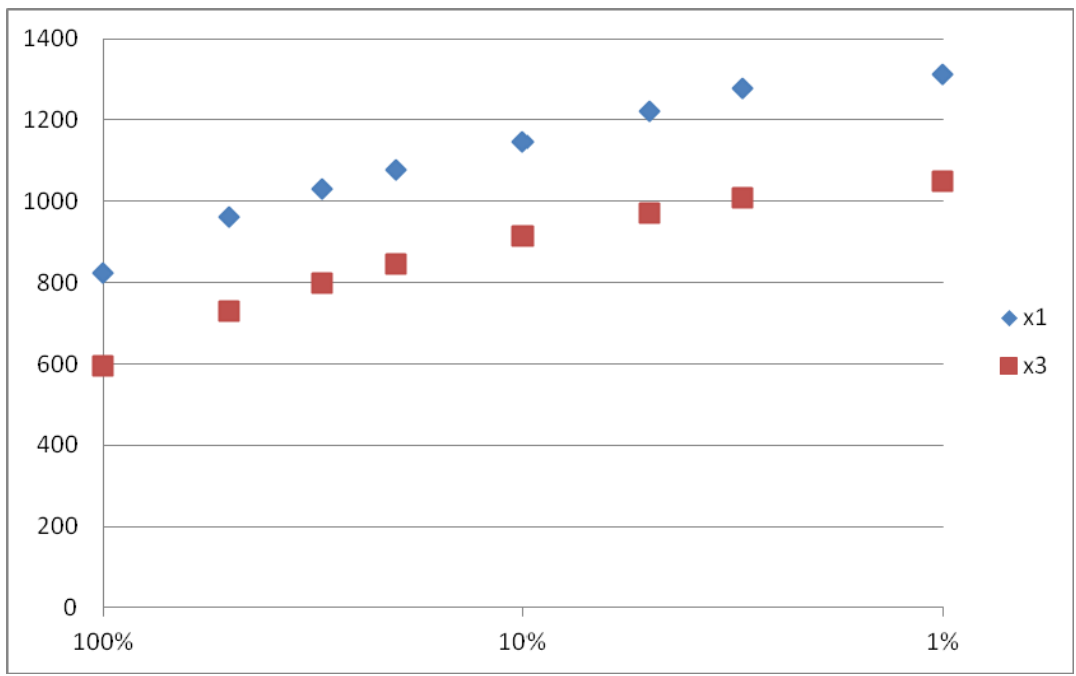

Figure 11. Sharpness measurements for different percentage selections on the frame sequence used in Section 5.1 taken on the NOT. The upper points are derived using conventional LI while the lower ones use the same quality/sharpness criteria for each frame but derive the position data by averaging the values for three frames (so averaging the derived positions from one frame before, the frame itself and one frame after). The frame rate is approximately 21 Hz so the lower curve corresponds to the frame rate equivalent to about 7 Hz.

In Fig. 11, we see the effect of averaging the reference star positions in order to improve signal-to-noise. The data are shown here use a bright reference star where there was no concern about reference star signal-to-noise in order to demonstrate the effect of averaging the coordinates used to centre frame by combining the values of the previous and the following frame with the frame itself. The quality selection criteria were still done on a frame-frame basis, only the coordinates measurements were averaged. We see that the LI method still works but at the cost of reducing the image sharpness. This is equivalent to reducing the frame rate to 7 Hz as far as the registration is concerned. In principle this result could be obtained with a reference star around one magnitude fainter than would normally be the limit.

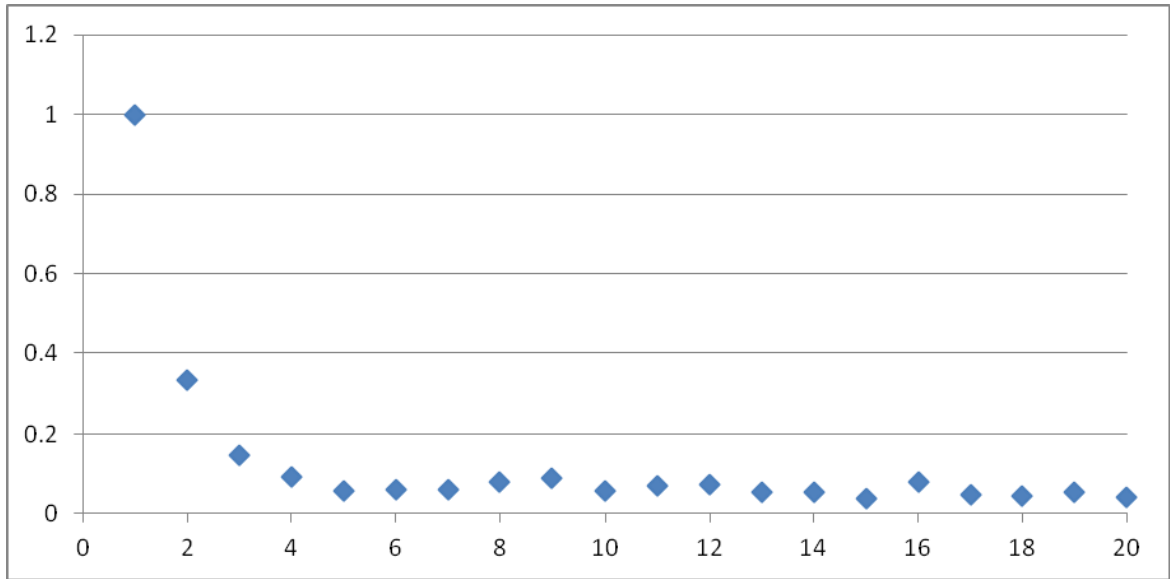

Figure 12. The autocorrelation function of the sharpness criteria used to rate the image quality as part of the LI procedure. It is clear that averaging this value would significantly compromise the LI procedure and indeed shows that ideally LI under these conditions should be run at a higher frame rate if at all possible.

Unfortunately averaging cannot be applied to the sharpness criteria applied to each frame of the sequence as part of the LI procedure. In Fig. 12, the rapid drop in the autocorrelation function of the sharpness criteria makes it clear that any reduction in the frame rate will have a significant negative effect on the capacity of LI to produce good images. Indeed the evidence is that ideally one should attempt to run at significantly faster frame rates than the 21 Hz used here.

## 7. CONCLUSIONS

Lucky Imaging is now established as an effective way of allowing ground-based telescopes to obtain near diffraction limited images in the visible. By the end of 2012 over 200 research papers had been published describing observations or results made by using LI, mostly on telescopes of diameters up to about 2.5 m. There is increasing interest in the use of LI in conjunction with a low order adaptive optics system to allow high resolution images to be obtained on larger telescopes and already the highest resolution images of faint targets have been obtained using this method. We have developed the method proposed and simulated by Garrel,

Guyon & Baudoz (2012) to combine conventional LI selection with selection in the Fourier plane so as to make best use of those images in which resolution is poorer in one direction and better in another. We have demonstrated that the improvement in sharpness, equivalent to Strehl ratio, delivered by these techniques is considerable. The resolution that had been achievable with 1% selection can now be delivered with 10% selection and the resolution that 10% selection produced with conventional LI can now be delivered with better than 50% selection using the LF methods. This is a considerable improvement in the observing efficiency that can be obtained with lucky imaging. As computing power continues to increase it will undoubtedly be possible to use these techniques routinely in future.

## ACKNOWLEDGEMENTS

The author would like to thank Vincent Garrel, Olivier Guyon and J Fienup for helpful conversations about these methods. He also is grateful to Olivier Guyon for many helpful suggestions that improved greatly the quality of the paper. He would also like to acknowledge the considerable help and support by the staff of the Nordic Optical Telescope in La Palma, and the staff of the California Institute of Technology at the Palomar Observatory who made these observations possible.